\documentclass[a4paper,aps,prl,reprint,superscriptaddress, twocolumn,preprintnumbers,amsmath,amssymb,nobalancelastpage,10pt, longbibliography]{revtex4-2}
\usepackage{svg}
\usepackage{amsmath}
\usepackage{verbatim}
\usepackage{graphicx}
\usepackage{color}
\usepackage{tikz}
\usepackage{upgreek}
\usepackage{float}

\usepackage[colorlinks=true,citecolor=blue,linkcolor=blue,urlcolor=blue]{hyperref}
\begin{document}

\title{Determination and correction of spectral phase from principal component analysis of coherent phonons}
\author{Emmanuel B. Amuah}
\affiliation{Department of Physics and Astronomy, Aarhus University, Ny Munkegade 120, 8000 Aarhus C, Denmark.}
\affiliation{ICFO - Institut de Ci\`encies Fot\`oniques, The Barcelona Institute of Science and Technology, 08860 Castelldefels, Barcelona, Spain}
\author{Khalid M. Siddiqui}
\author{Maurizio Monti}
\affiliation{Department of Physics and Astronomy, Aarhus University, Ny Munkegade 120, 8000 Aarhus C, Denmark.}
\author{Allan S. Johnson}
\affiliation{IMDEA Nanoscience, Calle Faraday 9, 28049, Madrid, Spain}
\author{Simon E. Wall}
\email[]{simon.wall@phys.au.dk}
\affiliation{Department of Physics and Astronomy, Aarhus University, Ny Munkegade 120, 8000 Aarhus C, Denmark.}

\begin{abstract}
Measuring the spectral phase of a pulse is key for performing wavelength resolved ultrafast measurements in the few
femtosecond regime. However, accurate measurements in real experimental conditions can be challenging. We show that the reflectivity change induced by coherent phonons in a quantum material can be used to infer the spectral phase of an optical probe pulse with few-femtosecond accuracy.
\end{abstract}
\maketitle

\section{Introduction}
The use of spectrally-resolved broadband pump-probe measurements is an increasingly popular method for understanding transient states in systems spanning from quantum materials\cite{Wall2010, Wall2012a, pastor_nonthermal_2022, novelli_witnessing_2014} to chemistry\cite{damrauer_femtosecond_1997, polli_conical_2010, cabanillas-gonzalez_pump-probe_2011}. By exploiting the full spectral dependence of the change in reflectivity or transmissivity of a sample, more detailed insights into the material response can be obtained by fitting the full transient response in the frequency domain.

A significant benefit of spectrally resolving measurements is that it does not require using a transform-limited probe pulse. Instead, the arrival time of the different colors (chirp) can be corrected post-hoc if the chirp can be determined, a procedure which has been shown both theoretically and experimentally to achieve transform-limited time-resolution~\cite{Kovalenko1999, polli_effective_2010}.

In many cases, the chirp can be corrected using information encoded
directly in the broadband pump-probe spectrogram. To correct the chirp,
typically the half-rise time of the material response is found as a
function of wavelength~\cite{Wegkamp2011a, cilento_ultrafast_2010, leblanc_temporal_2021, leblanc_phase-matching-free_2019}. However, determining this position depends on the function used to fit the rising edge, which is generally not known a priori. If multiple dynamics are occurring in the same spectra region, the rise-time may not accurately reflect the onset of the material response, and chirp correction will yield inaccurate results.

When the material dynamics are much slower than the rise time of the signal, small errors from determining the position of time zero,
t\textsubscript{0}, are not an issue. However, in many cases the early time dynamics are critical; for instance, in the field of quantum materials, as small delays of order tens of femtoseconds in the rising edge of the signal have been interpreted as evidence for light induced phase transitions~\cite{Cavalleri2004b, Polli2007}. Therefore, high-precision measurements of the spectral phase are essential. While full pulse characterization methods can provide this~\cite{johnson_measurement_2020, austin_spatio-temporal_2016}, such measurements cannot always be performed in-situ, which is critical given the extreme sensitivity to differences in material dispersion for the chirp of ultra-broadband pulses. Furthermore, probe pulses, which may be weak, may not have sufficient intensity for reliable measurements.

In this work, we show that the coherent phonon response of a quantum material, together with a principal component analysis of the transient change in reflectivity, can accurately determine spectral shifts in \emph{t\textsubscript{0}} with sub-fs precision. Such measurements can be performed with the same experimental setup as required for pump-probe measurements, i.e. in a cryostat and in a reflection geometry, and provide a robust, unambiguous method for determining \emph{t\textsubscript{0}} as a function of the probe wavelength.

\section{Coherent phonons and principal component analysis}

Coherent phonons are temporally coherent zone-center Raman-active phonons that are generated in solids when they are promptly excited by a pulse much shorter than the phonon period. In absorbing materials they are generated through the displacive mechanism in which the phonon oscillates with a cosine dependence about a displaced position~\cite{Zeiger1992}. Ignoring dephasing of the phonon and any electronic background effects, the wavelength-resolved transient change in reflectivity, \(\frac{\Delta R(\lambda,t)}{R}\), can be expressed as,

\begin{equation}
\frac{\Delta R(\lambda,t)}{R} = A(\lambda)\Theta\left( t - t_{0}(\lambda) \right)\left( 1 - \cos{\omega\left( t - t_{0}(\lambda) \right)} \right)  
\label{eq:1}
\end{equation}

where \(\Theta(t)\) is the Heaviside step function, \(t_{0}(\lambda)\) is the arrival time of the probe as a function of instantaneous
wavelength, and \(A(\lambda)\) is the modulation strength of the Raman active phonon of angular frequency $\omega$ on the reflectivity. Critically, from considerations of fundamental time-reversal symmetry, the temporal modulation due to the phonon must be frequency-locked and phase independent across the entire dielectric function, and thus any observed phase-shift with color is solely due to the arrival time of the spectral component, \(t_{0}(\lambda)\). In this sense the coherent phonons provide an absolute measure of the pulse chirp independent of any material properties or phase matching considerations.

In principle, one could fit the phonon modulation at each wavelength to correct for the chirp, with the accuracy of the chirp correction determined by the accuracy in the fit. However, this approach is computationally expensive and can become prohibitively sensitive to noise. Instead, we use principal component analysis (PCA) to reduce the dimensionality of the problem and directly access the phonon modulation.

PCA takes a given muti-dimensional function and breaks it down into a linear summation of ``eigen'' functions (principal components) of each variable, i.e. 
\begin{equation}
g(\lambda,\ t) = \sum_{i}^{}{\Lambda_{i}(\lambda)T_{i}(t)} 
\label{eq:2}
\end{equation}

To understand what to expect from a PCA analysis of a coherent phonon signal, we re-write equation~\ref{eq:1} in the form of equation 2. First, we note that the Heaviside step function is not separable and thus can be expressed as an infinite series in the same form as equation~\ref{eq:2}

\begin{equation}
  \Theta\left( t - t_{0}(\lambda) \right) = \sum_{i}\Omega_{i}(\lambda)\Psi_{i}(t)  
  \label{eq:3}
\end{equation}

Whereas the cosine term can be expressed with just two terms

\begin{multline}
  \cos \left(\omega\left(t-t_{0}\left(\lambda\right)\right)\right) = \cos\left(\omega t\right)\cos\left( \omega t_0\left( \lambda \right) \right) \\ + \sin \left(\omega t \right)\sin\left(\omega t_0\left(\lambda \right) \right)
  \label{eq:4}
\end{multline}

Therefore, we expect a PCA analysis of chirped transient reflectivity signals to return a pair of PCs corresponding to each of the linear contributions in equation~\ref{eq:3}, which will have temporal factors oscillating in and out of phase. The smaller the chirp, the weaker the second component. The ratio of the amplitude of these pairs is then simply \(\tan\left( \omega t_{0}(\lambda) \right)\). Thus, a high frequency phonon can be used to determine the phase shift precisely.

\begin{enumerate}
\def\labelenumi{\arabic{enumi}.}
\setcounter{enumi}{2}
\item
  Application of PCA to experimental data
\end{enumerate}

To demonstrate this principle, we perform transient reflectivity on the manganite Pr\textsubscript{0.5}Ca\textsubscript{1.5}MnO\textsubscript{4} (PCMO), a quantum material that exhibits charge and orbital ordering below 320 K, with a 16 THz (533.6 cm\textsuperscript{-1}) Raman active Jahn-Teller mode~\cite{Handayani2015}, which we exploit here.

\begin{figure}
    \centering
    \includegraphics[width=8.6cm]{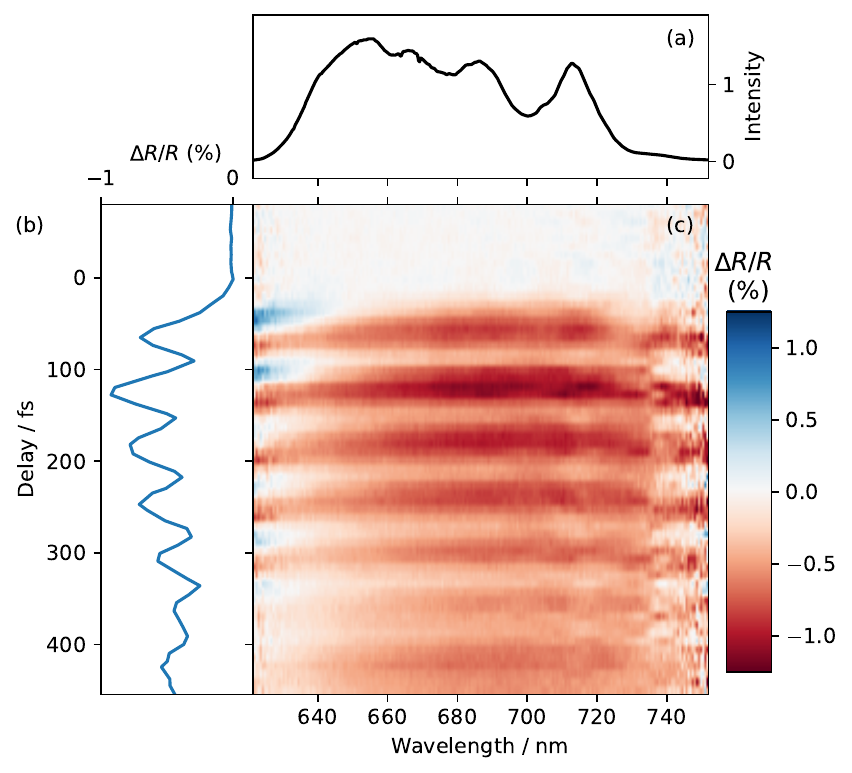}
    \caption{(a) Measured probe spectra reflected from PCMO at 120 K without pump. (b) Spectrally integrated transient reflectivity following excitation. (c) Wavelength resolved change in reflectivity.}
    \label{fig:1}
\end{figure}

We excite the system with 12 fs pulses centered at 1.78 µm and probe the system using pulses from a commercial Non-colinear Optical Parametric Amplifier (NOPA) as described in Ref.~\cite{amuah_achromatic_2021}. The pump fluence is set to 1 mJ/cm2 to be in the linear regime and the reflected light is captured on a CCD spectrometer. The sample is mounted inside an optical cryostat and cooled to 120 K to reduce the effects of damping on the phonon mode and maximize the amplitude. The transient reflectivity is shown in Figure~\ref{fig:1}. The 16 THz mode, together with an electronic background can be clearly seen over the whole wavelength range.

\begin{figure}
    \centering
    \includegraphics[width=8.6cm]{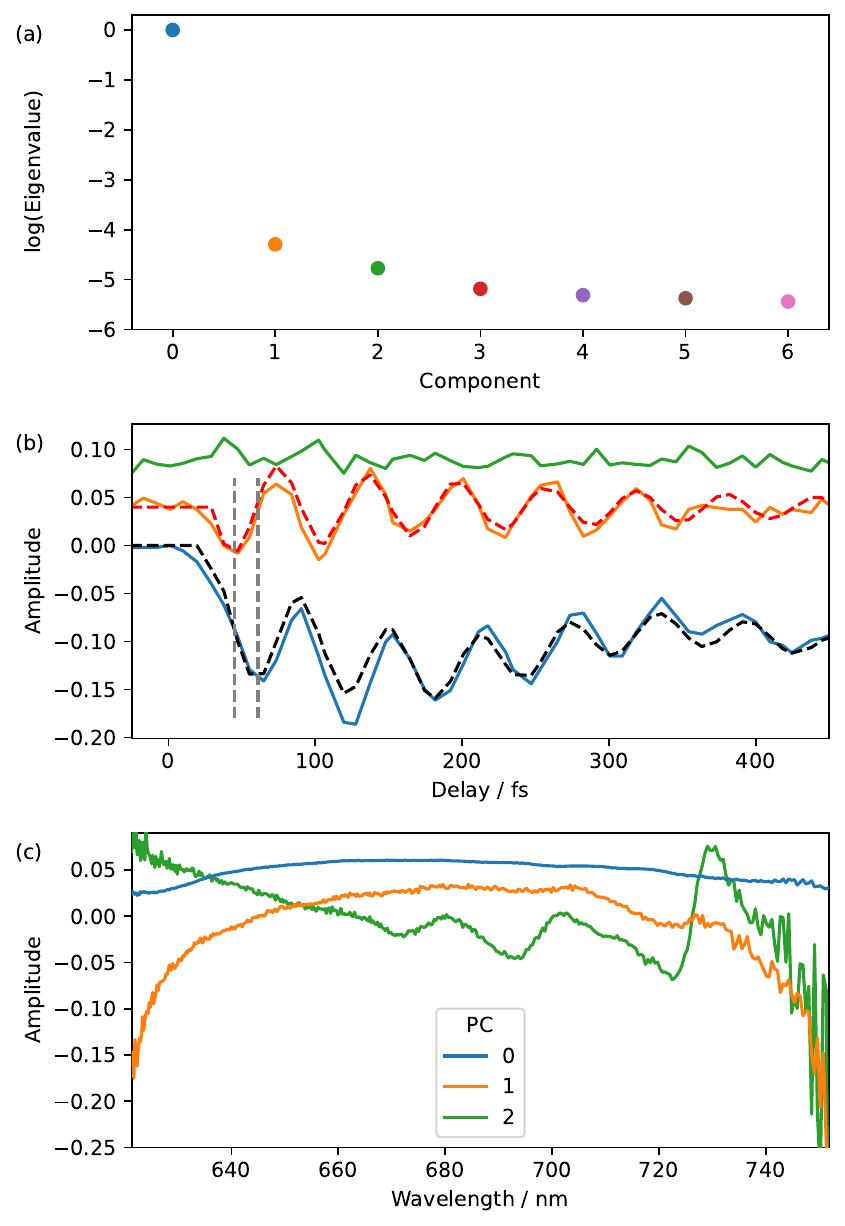}
    \caption{(a) Logarithm of the normalized eigenvalue of each principal component. The majority of the transient signal can be captured by PC0. (b) Time traces associated with each PC. PC0 and PC1 show the out-of-phase relationship of the 16 THz mode expected for a chirped probe, indicated by the dashed vertical lines. PC2 (green line) does not show any meaningful dynamic, indicating subsequent terms of the PCA are below the noise. Dashed lines are the fits used to extract the amplitude of the 16 THz phonon (traces offset for clarity). (c) Wavelength dependence of the first three PCs.}
    \label{fig:2}
\end{figure}

The results of a PCA analysis on this data are shown in Figure~\ref{fig:2}. Fig.~\ref{fig:2}a shows the eigenvalues of the PCA, which indicates the relative contribution of each component, normalized to the strength of the dominant component (PC0). Most of the signal is captured by a single component, PC0. The dynamics of PC0 are plotted in Fig.~\ref{fig:2}b, which show that the main component closely follows the integrated response shown in Fig.~\ref{fig:1}b. This is expected if the dynamics are independent of wavelength
and the pulse is compressed. The second component (PC1) shows the out-of-phase oscillations expected for a chirped probe, as highlighted by the vertical dashed lines. Fig.~\ref{fig:2}c shows the spectral amplitude of PC1 is most significant at the edges of the spectra, further supporting its assignment as a measure of the chirp, as this matches expectations for a broadband pulse which has traversed significant material. Also shown in Fig~\ref{fig:2}b and c is the time trace for PC2, which shows no significant
dynamic above the noise, indicating that subsequent terms in the PCA do not add any meaningful contributions to the signal.

To determine the magnitude of the chirp we need to extract the amplitude of the oscillation from the oscillation from each PC. Additional terms beyond those in Eq.~\ref{eq:1} are needed to fit the dynamics, most clearly seen in the signal not oscillating around zero. This comes from other degrees of freedom in the system. such as electronic relaxation and a second 3.1 THz phonon mode. Therefore, we fit PC0 by a sum of exponentials and two phonon terms oscillating with a cosine-like phase, representing the 3 THz and 16 THz mode. We simultaneously fit PC1 with the same 16 THz oscillator, but this time with a sine-like dependence.
In principle other terms could appear in PC1, or in other PCs, but we found the data could be well fit by only including the sine term in PC1. An excellent fit can be achieved for both datasets with a retrieved phonon frequency of 16.4 THz and a common phase offset.

Having fit both significant PCs, we then multiply the wavelength component of each PC (Fig.~\ref{fig:2}c) by the phonon amplitude extracted from the fit of the corresponding temporal PC (Fig.~\ref{fig:2}b). The ratio of these two terms gives the tangent of phase as a function of wavelength, $\psi$. This can be converted into a time shift via \(t_{0} = \phi(\lambda)/\omega\), where \(\omega\) is the angular frequency of the 16.4 THz mode. Importantly, this does not depend on the recovered phase offset or exponential functions, which correspond to the conventional method of fitting the rise-time and are model-dependent.

\begin{figure}
    \centering
    \includegraphics[width=8.6cm]{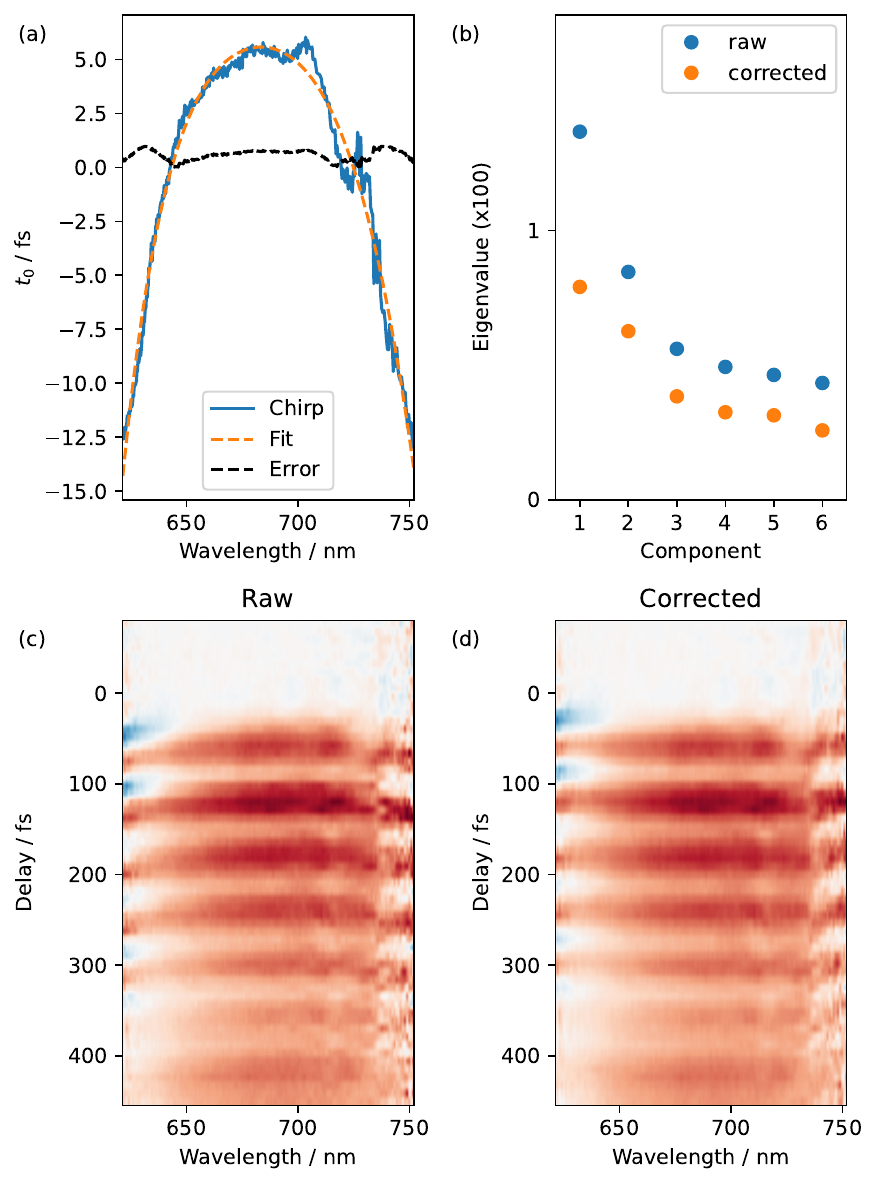}
    \caption{(a) Extracted pulse arrival time (phase) as function of wavelength (solid blue), polynomial fit to the chirp (dashed orange), and associated error (dashed black line). (b) Change in normalized amplitude of PCs following chirp correction (linear scale), PC0=1 is not show. The amplitude of all high-order PCs is significantly reduced, showing the improved separability of spectral and temporal degrees of freedom. (c) and (d) Raw and corrected pump-probe traces showing clear improvement following chirp correction. The color scale is the same as in Figure~\ref{fig:1}.}
    \label{fig:3}
\end{figure}

Figure~\ref{fig:3} shows the extracted arrival time as a function of probe wavelength, which corresponds to a pulse with a group delay of just 0.4 fs\textsuperscript{2} and third order dispersion of -943 fs\textsuperscript{3}. The error in the arrival time is determined from the accuracy of global fit parameters, specifically the amplitude of the phonon mode for the two components and the determination of the frequency. We approximate this error to be 10\% of the best fit value, which is slightly above the value obtained based on variance of the fitting. The error in determining time zero is shown by the dashed line in Fig.~\ref{fig:3}a. A maximum error of 1 fs is found at the edges of the spectrum, and is approximately a factor of two lower in the center. This is significantly below the phonon period and superior to what can be achieved by fitting individual wavelengths. A maximum delay of 18 fs is recorded across the entire bandwidth. To check the accuracy of this chirp, we then time shift each wavelength by the corresponding delay and repeat the PCA on the corrected data set. The result is shown in Fig.~\ref{fig:3}b, which shows a significantly suppressed amplitude for PC1. Furthermore, the improvement can be directly seen when comparing the raw wavelength-resolved data (Fig.~\ref{fig:3}c) to the corrected plot (Fig.~\ref{fig:3}d).

\begin{figure}
    \centering
    \includegraphics[width=8.6cm]{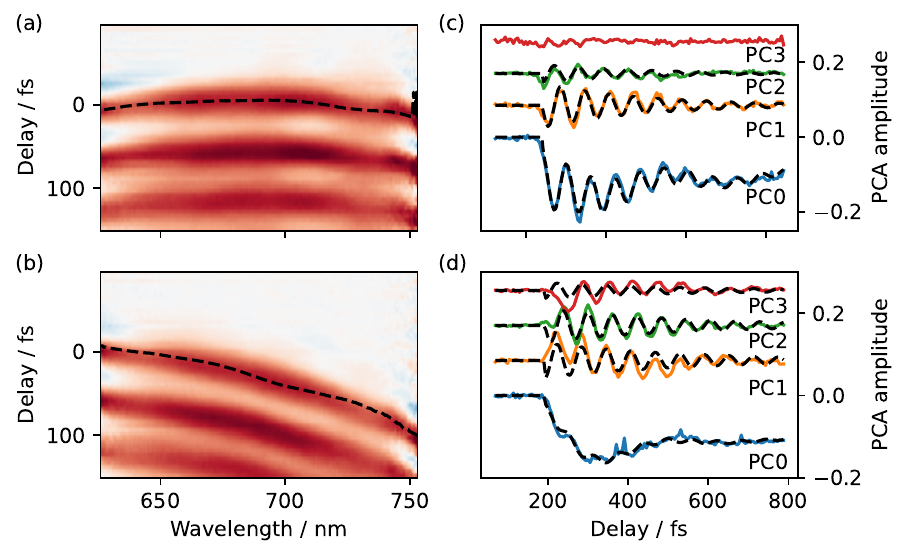}
    \caption{A zoom of the transient reflectivity around \emph{t\textsubscript{0}} to highlight the dispersion of pulse (a) a nearly compressed pulse and (b) probe pulse passes through 5.3 mm of fused silica. An offset of 8099 fs has been applied to the time zero position in (b) when compared with (a) to account for the delay induced by the material. (c) PCA decomposition of (a) showing that there are three transient components. (d) PCA decomposition of (d) showing that the 4\textsuperscript{th} PC becomes significant (PC4). Dashed lines in (c/d) correspond to fits as discussed in Fig.\ref{fig:2}. The dashed lines in (a/b) correspond to the retrieved phase. The color scale for (a) and (b) is the same as Figure~\ref{fig:1}.}
    \label{fig:4}
\end{figure}

To further check the validity of this approach, we test how well our procedure works when a known amount of glass is added to probe-path. In the following, the laser was in a different condition to the previous results resulting in a slightly different pump duration and probe spectrum. We introduced a 5.3 mm thick window of fused silica into the probe beam and compared the phase shift retrieved using our method to that expected based on literature values of the refractive index of the additional material~\cite{malitson_interspecimen_1965, lee_measurement_2016}. Figure~\ref{fig:4} shows the transient reflectivity before (a) and after (b) the addition of the fused silica plate. A strong chirp is clearly introduced into the beam by the additional glass. In this experiment, the PC analysis results in more PC components (Figs \ref{fig:4}c and d), even in the compressed case. These terms arise due a combination of slightly different pump energy and duration, inducing a stronger \(\Theta\left( t - t_{0}(\lambda) \right)\) term, which introduces more oscillating pairs. However, we can still extract the phase by summing the amplitudes of the modes with the same phase.

The extracted phase is then shown as the dashed lines in Fig.\ref{fig:4}a and b. Despite the increase in PCs needed to represent the signal, the extracted phase closely tracks the peak of the oscillation demonstrating that the chirp is faithfully determined by the ratio of the PC components.

\begin{figure}
    \centering
    \includegraphics[width=8.6cm]{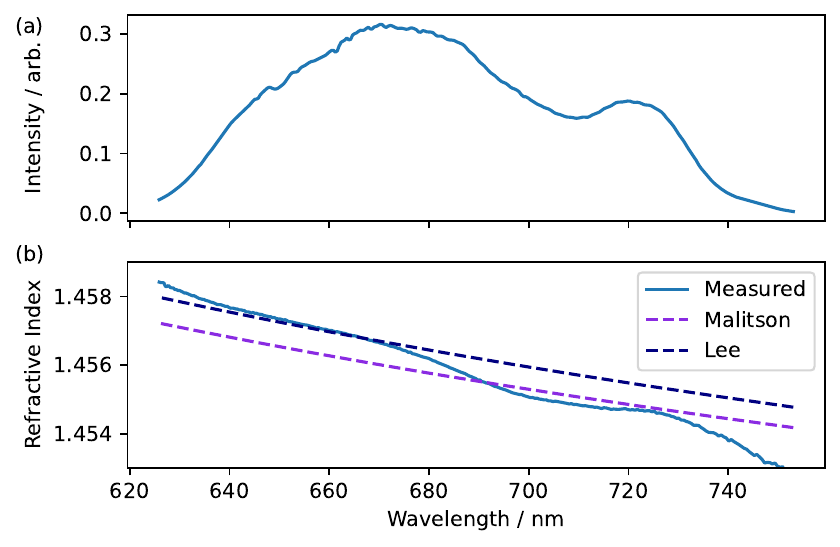}
    \caption{(a) Spectrum of the probe pulse used in Figure~\ref{fig:4}. (b) Calculated refractive index based on the delay determined from the PCA (solid line) compared to the results of Malitson~\cite{malitson_interspecimen_1965} and Lee et al.~\cite{lee_measurement_2016}. The results obtained with our method lie within the reported values over the region in which our spectrum is most intense.}
    \label{fig:5}
\end{figure}

Figure~\ref{fig:5} shows the probe spectrum used in figure~\ref{fig:4}, and the refractive index obtained from the PCA. The refractive index, \emph{n}, is obtained from
\begin{equation}
n = 1 + \frac{c\Delta t(\lambda)}{d},    
\end{equation}
where \emph{c} is the speed of light, \(\Delta t(\lambda)\) is the difference in the extracted delay as a function of vacuum wavelength
with and without the SiO\textsubscript{2} block, and \(d\) is the sample thickness. The extracted refractive index is within the range of values that have been reported in the literature, generally agreeing with tabulated values down the 0.1\% level~\cite{malitson_interspecimen_1965, lee_measurement_2016}, demonstrating the effectiveness of our approach. The majority of the discrepancy can be thus be put down to variations in the refractive index from different samples due to dopants or other defects.

\section{Discussion}  
Our approach, using a known transient response to track the synchronization of the different spectral components in a pulse but
using PCs to reduce the dimensionality of the problem, dramatically reduces the number of fit parameters and noise sensitivity compared to fitting at each wavelength independently. The ensuing gains in signal-to-noise and reduced fit parameters will in general depend on the degree of dimensionality reduction possible. Our approach is also much more robust to complex chirp profiles than existing methods, as the factorization procedure does not depend explicitly on the chirp. For instance, randomizing the spectral columns in the data set shown here, which would correspond to a very complex spectral chirp and a white-noise-like pulse in a real system, does not affect the linear decomposition of PCA, and the chirp could be equally well corrected. Furthermore, the technique can easily be applied in-situ with most transient absorption setups, does not suffer from phase-matching limitations, is unambiguous due to the fundamental frequency and phase locking of the modulation to the phonon, and thus could be ideally suited for characterizing ultra-broadband supercontinuua. Finally, while here we use cryogenically cooled PCMO as our target, there are likely more optical choices that can improve the accuracy. The main factor that can improve the accuracy is the phonon frequency, materials with a high frequency and weakly damped phonons will give the best results. Materials with a strong step-like transient or multiple phonon modes will need more PC terms to describe, this increase in fitting complexity will reduce the accuracy, but we point out that slow modes will experience a smaller phase shift for a given chirp and so will not induce large sine-like terms and will not have a big impact. This suggests that diamond could be a good choice~\cite{ishioka_coherent_2006} for determining the phase with a high accuracy.

In summary, we have demonstrated a novel in-situ method for post-hoc chirp correction in broadband pump-probe measurements using the PCA of coherent phonon modulations. The accuracy of the method is determined by quality of the fitting and the frequency of the phonon modulations. Further improvements could be obtained with materials that exhibit even higher-frequency phonons and longer dephasing times, though the necessity of resolving the oscillations places an upper bound on the phonon frequency that can be used depending on the transform-limited pulse duration of the probe. The robustness and simplicity of our method, which uses only information already encoded in many spectrally-resolved pump-probe traces, makes it ideal for many applications in ultrafast spectroscopy where existing methods may struggle.

\begin{acknowledgements}
\textbf{Acknowledgements.} The authors thank Chris Brahms for several stimulating discussions.
\end{acknowledgements}

\bibliography{Bib.bib}

\end{document}